\documentclass[oldversion]{aa}

\usepackage{epsfig}
\usepackage{graphics}
\usepackage{float}
\usepackage{amsmath}
\usepackage{multirow}
\usepackage{longtable}
\usepackage{rotate}
\usepackage{array}
\usepackage{subfigure}
\def\kms{\ifmmode{\rm km\,s^{-1}}\else\hbox{$\rm km\,s^{-1}$}\fi}

\setlongtables

\begin{document}

\title{Coronal winds powered by radiative driving}

\author{L.B.Lucy}
\offprints{L.B.Lucy}
\institute{Astrophysics Group, Blackett Laboratory, Imperial College 
London, Prince Consort Road, London SW7 2AZ, UK}
\date{Received ; Accepted }

\abstract{A two-component phenomenological model developed originally
for $\zeta$ Puppis is revised in order to model the outflows of
late-type O dwarfs that exhibit the {\em weak-wind phenomenon}.
With the theory's standard parameters for a generic weak-wind star, the
ambient gas is heated to coronal temperatures $ \approx 3 \times 10^{6}$K
at radii $ \ga 1.4 R$, with cool radiatively-driven gas being then 
confined to dense clumps with filling factor $\approx 0.02$. 
Radiative driving ceases at radius $ \approx 2.1R$ when the clumps are
finally destroyed by heat conduction from the coronal gas.  
Thereafter, the outflow is
a pure coronal wind, which cools and decelerates reaching $\infty$
with terminal velocity $\approx 980$ km s$^{-1}$.   
}

\authorrunning{Lucy}
\titlerunning{Coronal winds}
\maketitle

\section{Introduction}

The X-ray emission from O stars (Harnden et al. 1979) is now generally agreed
to arise from numerous shock fronts distributed throughout their winds.
An early theory of such X-ray emitting winds  
(Lucy \& White 1980; LW) was based on a two-component phenomenological
model for the finite amplitude state reached by unstable line-driven winds. 
Subsequently, the fundamental
approach of computing the growth of the instability using the 
equations of radiation gas dynamics was pioneered by 
Owocki et al. (1988) and Feldmeier (1995), albeit with the then necessary
restrictions to 1-D flow and simplified radiative transfer.

A question meriting further research is how and where this wind-shock
model fails. 
According to LW, failure occurs at the low 
mass-loss rate ($\Phi$) of a main sequence B0 star, because the assumption 
of rapid radiative cooling of shocked ambient gas then
breaks down for blob velocities $ v_{b} \ga 10^{3}$km s$^{-1}$, 
resulting in the heating of the blobs and consequent loss of line-driving.
They conjecture that 'thereafter, the relative motions of the two components 
dissipate and the smoothed wind coasts out to infinity.' In effect, LW
suggest that a wind that is initially radiatively driven converts
into one that relies on thermal pressure to reach $\infty$ - i.e., 
a coronal wind.     

In addition to this question's intrinsic interest, it is notable that 
the locus of this expected failure coincides  
with that of stars exhibiting the {\em weak-wind phenomenon}.
(e.g., Marcolino et al. 2009; M09). 
Accordingly, this paper elaborates LW's conjectures for the
outflow from a weak-wind star.

\section{A Multi-zone wind: Zone 1}

The two-component model must be generalized to remove the
assumption of instantaneous cooling of shocked gas and to incorporate
blob destruction at finite radius. To
achieve these aims, a multizone model is adopted, with each zone corresponding
to different physical circumstances. 

Zone 1 starts just beyond the sonic point and ends when the isothermal-shock
assumption is no longer justified. 
We assume instability has grown to full amplitide
and adopt the LW description in which radiative-driven blobs ($b$) interact 
dynamically with a low density ambient ($a$) medium. Apart from the 
infinitesimally-thin
cooling zones at shock
fronts, both components are in thermal equilibrium with the
photospheric radiation field, so that $T_{a,b} = T_{eq}$.
 
\subsection{Blobs} 

The blobs can be identified with the clumps that are now a standard and
spectroscopically-required feature of diagnostic codes for
O-star winds (e.g., Bouret et al. 2005). In such codes,
the clumps are assumed to obey the $\beta-$velocity law
\begin{equation}
  v_{b} = v_{\infty} \left( 1 - \frac{R}{r} \right)^{\beta}
\end{equation}
where $R$ is the photospheric radius and $v_{\infty}$, the terminal velocity,
is determined from the violet edges of P Cygni absorption troughs.

Given the wide use of the $\beta$-law, this now replaces
LW's Eq. (10). 
But here, since Eq. (1) ceases to apply when $r > r_{S}$, the blobs' 
destruction radius, $v_{\infty}$ is not an observable.
The highest velocity at which UV absorption is detected is a measure  
not of $v_{\infty}$ but of $v_{b}(r_{S})$.    

For given $\beta$, diagnostic modellers choose the clumps' filling factor
$f_{b}$ and mass-loss rate $\Phi_{b}$ so that absorption troughs have their
observed strengths. For a strong line at frequency $\nu_{0}$, this
typically requires that, despite clumpiness, a continuum photon emitted 
between $\nu_{0}$ and $\nu_{0}(1+v_{\infty}/c)$ has small probability of
escaping to $\infty$, and so most of the photon momentum in 
this interval is transferred to the clumps. In an LW wind, essentially 
the same requirement arises as a consistency criterion: 
since the ambient gas is assumed {\em not} to be radiatively driven, it must be 
shadowed by the blobs. 
The optical depth criterion adopted by 
LW is that $\tau_{1}(r) > 1.5$ for all $r$, where $\tau_{1}$ is given in 
Eq.(11) of LW. For
the $\beta-$velocity law, the function $\chi$ in their $\tau_{1}$ formula 
becomes
\begin{equation}
   \chi(x;\beta) = \frac{1}{2} \: (1-x)^{1-2 \beta} \:/\:
                                     [2 \beta + x - (1+\beta)x^{2}]
\end{equation}
where $x = R/r$.

\subsection{Dynamics} 

In zone 1, the dynamical interaction of the blob and ambient  
components is treated exactly as in LW. For more recent treatments and 
applications, see Howk et al. (2000) and Guo (2010). 

The equation of motion obeyed by the blobs is
\begin{equation}
   v_{b} \: \frac{d v_{b}}{d r} \: = \: g_{R} - g_{D} - g
\end{equation}
where $g_{R}, g_{D}$, and $g$ are the forces per gram due to radiation, drag, 
and gravity, respectively. 
The drag force $g_{D}$ retards the blobs but accelerates the ambient gas
and is the means by which photon momentum is transferred to this component.
The resulting equation of motion of the ambient gas is 
\begin{equation}
  v_{a} \: \frac{d v_{a}}{d r} \: = \: 
   -\frac{1}{\bar{\rho}_{a}}\frac{d \bar{P}_{a}}{d r} + 
   \frac{\bar{\rho}_{b}}{\bar{\rho}_{a}  } \: g_{D}- \: g
\end{equation}
Here $\bar{\rho}_{a,b}$ are the smoothed densities, and 
$\bar{P}_{a} = a_{a}^{2} \bar{\rho}_{a}$ with $a_{a}^{2} = kT_{a}/\mu m_{H}$. 

With the LW assumption of no mass exchange between the components,
the two equations of continuity integrate to give
\begin{equation}
   \Phi_{a,b} = 4 \pi r^{2} \bar{\rho}_{a,b} v_{a,b} 
\end{equation}
where $\Phi_{a,b}$ are constants whose sum is the star's mass-loss rate
$\Phi$.

The drag force $m g_{D}$ on a blob of mass $m$ is computed using De Young and
Axford's (1967) theory of inertially-confined plasma clouds - see
Sect. II b) in LW. The
resulting formula is
\begin{equation}
   m g_{D} \: = \: \frac{1}{2} \: C_{D} \rho_{a} U^{2} {\cal A}_{b} 
\end{equation}
where ${\cal A}_{b} = \pi \sigma^{2}$ is the blob's mean cross section, 
$U = v_{b} - v_{a}$ is the blob's velocity relative to the ambient gas,
and the drag coefficient $C_{D} = 1.519$.

\subsection{Filling factors} 

At a point $(v_{b},v_{a},r)$ in an outward integration 
with specified $\Phi_{a,b}$,
the smoothed densities $\bar{\rho}_{a,b}$ are given by Eq. (5). The ambient
density is then $\rho_{a} = \bar{\rho}_{a}/f_{a}$, where $f_{a}$ is the filling
factor of the ambient gas. Correspondingly, the mean density  
of the stratified De Young - Axford blob is 
$\rho_{b}= \bar{\rho}_{b}/f_{b}$. Now, in the absence of a void 
component, $f_{a} + f_{b} = 1$, and so only one of $f_{a}$ and $f_{b}$ is 
independent.
To determine $f_{b}$, say, we must iterate.   
Solution by repeated bisection is adopted, starting with 
upper and lower limits $f_{U} = 1$ and $f_{L} = 0$. Then, with the estimate 
$\tilde{f_{b}} = (f_{U}+f_{L})/2$, the blob's volume $V_{b}$ is computed from
LW's Eq.(5). The mean density of the blob is then  
$\rho_{b}= m/V_{b}$, corresponding to $f_{b} = \bar{\rho}_{b}/\rho_{b}$.
If $f_{b} < \tilde{f_{b}}$, the new upper limit is $f_{U} = \tilde{f_{b}}$. 
On the other hand, 
if $f_{b} \geq \tilde{f_{b}}$, the new lower limit is $f_{L} = \tilde{f_{b}}$. 
The iterations continue until $f_{U} - f_{L} < 10^{-7}$. Then, 
with the resulting converged value of $f_{b}$, all quantities required to 
continue the integration can be evaluated.    

\subsection{Switch criterion} 

The assumption of instantaneous cooling breaks down
at low densities because the cooling rate per unit volume
$\dot{{\cal C}} \propto \rho^{2}$.
If the cooling time scale $t_{c}$ increases to the extent that
a parcel of shock-heated gas encounters
another shock before cooling back to $T_{eq}$, then shock-heating 
raises the mean temperature of the ambient medium. 
An approximate criterion for this transition to zone 2 is derived as follows:

First, consider radiatively-cooled flow of monatomic gas ($\gamma = 5/3$)
emerging from a 
steady shock (see Fig.1 in Draine \& McKee 1993). Since this 
flow is subsonic, the pressure gradient may be neglected in comparison to that
of temperature. Thus, in the shock's frame,
\begin{equation}
   \frac{d \ln T}{dr} \: \approx \: - \frac{2}{5} \: 
             \frac{\dot{{\cal C}}}{Pv} \: = \: \frac{1}{\ell_{c}}
\end{equation}
The cooling timescale is therefore 
$t_{c} = \ell_{c}/v = 5/2 \times P/\dot{{\cal C}}$.

Now consider flow into the bow shocks. The entire mass $\rho_{a}$ of ambient 
gas in unit volume is shocked
in time interval $t_{i}= \rho_{a}/{\cal N}_{b} j_{b}$, where
${\cal N}_{b} = f_{b}/V_{b}$ is the number density of blobs, and 
$j_{b} \approx \rho_{a} U {\cal A}_{b}$ is the mass flow rate through each
bow shock. Hence $t_{i} \approx 4/3f_{b}  \times \sigma /U$.
  
The criterion for switching from zone 1 to zone 2 is then simply
$t_{c} > t_{i}$.

\section{A Multi-zone wind: Zone 2}

The outward integration of the wind continues in zone 2 with 
the same basic model except that
$T_{a,b} \ne T_{eq}$. The ambient gas is now heated by 
being repeatedly shocked, and the blobs in turn gain heat by 
conduction from the ambient gas. Zone 2 ends when the blobs can no longer
achieve thermal equilibrium.

\subsection{Blob survival}

The survival of blobs (clumps) in stellar winds
has similarities to that of clouds in the interstellar medium.
In that context, Cowie \& McKee (1977) studied the
evaporation of a spherical cloud embedded in a hot tenuous medium. 
Importantly, they treated the saturation of heat conduction  
when the electron mean free path in the surrounding medium is $\ga$ the 
cloud's radius and estimated, under the assumption of steady outflow,
the reduced evaporation rate.
In a companion paper (McKee \& Cowie 1977; see also Graham \& Langer 1973),   
they consider the effects of radiative losses, finding that
evaporation is replaced by condensation if the losses exceed the 
heat input from the hot gas. However, numerical calculations by 
Vieser \& Hensler (2008) cast doubt 
on the assumption of steady outflow. In the case of saturated conduction,
they find a further reduction of evaporation rate by a factor $\sim 40$ 
due to changes of the cloud's environment caused by the outflow. 

Given the evident difficulty of reliably predicting 
when cool gas is eliminated by its interaction
with surrounding hot gas, a simple prescriptive approach is adopted here:
As in zone 1, the blobs retain their fixed mass $m$ throughout
zone 2. However, when the heat input from the ambient gas exceeds
their maximum cooling rate, the blobs are assumed to merge
{\em instantly} with the ambient gas.   

\subsection{Heating and cooling of blobs}

In zone 2, the blobs are surrounded by shock-heated gas and so will
be heated by thermal conduction. But if the ambient gas reaches 
coronal temperatures, heat conduction is flux-limited. 
Moreover, conductivity may be suppressed by
magnetic fields. An approximate formula interpolating 
between the classical and saturated limits and incorporating a suppression
factor $\phi$ is derived in Appendix A.   

If ${\cal L}_{in}$ is the rate of heat flow from the ambient gas,
a blob will achieve thermal equilibrium at $T_{b} > T_{eq}$ if the
{\em enhanced} radiative cooling rate   
\begin{equation}
   \Delta {\cal L}_{b}  \: = \: (n_{e} n_{H})_{b} \: 
                     (\Lambda(T_{b}) - \Lambda(T_{eq}))   \times V_{b}
                  \:  =  \: {\cal L}_{in} 
\end{equation}
Here $\Lambda(T)$ is the optically-thin cooling function, and the blob is
treated as isothermal and of uniform density. (But note that LW's definition
of $\rho_{b}$ is such that $\Delta {\cal L}_{b}$ is exact for the density
stratification of an isothermal De Young-Axford blob.) 

Because $\Lambda(T)$ reaches a maximum at $T_{\dag}(K) = 5.35$ dex 
(Dere et al. 2009), the solution of Eq.(8) with $T_{b} < T_{\dag}$ is
appropriate as the blobs are heated to above $T_{eq}$. When $T_{b}$ 
reaches $T_{\dag}$, the corresponding ${\cal L}_{in}$ is the maximum value
consistent with thermal equilibrium. Any further increase in  ${\cal L}_{in}$
cannot be matched by increased cooling. Accordingly, we
take this as the point beyond which the blobs cannot survive.     

Note that if conduction is completely suppressed $(\phi =0)$, then
${\cal L}_{in} = 0$ and the solution of Eq.(8) is $T_{b} = T_{eq}$.
The blobs therefore survive, and zone 2 extends to $\infty$.

\subsection{Heating and cooling of ambient gas}

According to LW, the rate at which energy is being dissipated per unit volume
is 
\begin{equation}
   \dot{Q} \: = \: g_{D} \: \bar{\rho}_{b} \: U
\end{equation}
Dividing by ${\cal N}_{b}$, we find that the rate per blob is
\begin{equation}
   \dot{Q}_{b} \: = \frac{1}{2} \rho_{a} U^{2} \times U \times C_{D} {\cal A}_{b
}
\end{equation}
showing that in unit time each blob's bow shock dissipates the kinetic energy 
in a column of inflowing gas of length $U$ and 
cross section $C_{D} {\cal A}_{b}$. In zone 1, this dissipated energy is 
radiated by a thin cooling layer, and
so  $\dot{Q}$ determines the sum of these layers' frequency-integrated
emissivities - Eq.(7) in LW. But in zone 2 where $t_{c} > t_{i}$, we jump to
the opposite limit, treating dissipation as a heat source distributed
uniformly throughout the ambient gas, and similarly for cooling.
Accordingly, the energy equation for stationary flow of the monatomic
ambient gas is 
\begin{equation}
   v_{a} \left( \frac{d P_{a}}{d r} - \frac{5}{3} \: a_{a}^{2}
            \frac{d \rho_{a}}{d r} \right) \: = \:
                  \frac{2}{3} \: (\dot{Q} -  \dot{\cal C}_{tot})
\end{equation}
Note that since $f_{a} \approx 1$ terms arising from radial changes 
in  $f_{a}$ have been neglected.

The total cooling rate per unit volume of ambient gas, $\dot{\cal C}_{tot}$,
is the sum of the losses due to radiative cooling and to
conduction into the blobs. Thus,
\begin{equation}
   \dot{\cal C}_{tot} \: = \:  (n_{e} n_{H})_{a} \:  \Lambda(T_{a})
                 + {\cal N}_{b} \; {\cal L}_{in} 
\end{equation}
Integration of Eqs. (4),(5) and (11)
continues until $T_{b} = T_{\dag}$, at which point the blobs are deemed
to merge instantly with the ambient gas (Sect. 3.1). Accordingly, 
the transition from zone 2 to zone 3 occurs at $S$, a surface of discontinuity 
(e.g., Landau \& Lifshitz 1959), across which 
the fluxes of mass
\begin{equation}
  J = J_{a} + J_{b} = f_{a} \rho_{a} v_{a} + f_{b} \rho_{b} v_{b}
\end{equation}
momentum
\begin{equation}
 \Pi = f_{a} (P_{a} +  \rho_{a} v_{a}^{2}) + 
                             f_{b} (P_{b} +  \rho_{b} v_{b}^{2})
\end{equation}
and energy
\begin{equation}
 F = J_{a} \: \left( \frac{1}{2} v_{a}^{2} + \frac{5}{2} a_{a}^{2}   \right)  + 
        J_{b} \:\left( \frac{1}{2} v_{b}^{2} + \frac{5}{2} a_{b}^{2}   \right)  
\end{equation}
are continuous. Zone 2 thus ends at $r_{S}$ with the evaluation of
$J, \: \Pi$ and $F$.

\section{A Multi-zone wind: Zone 3}

The outward integration continues in zone 3, but now the blobs have
disappeared, leaving a single fluid component with no driving force 
$(g_{D} = 0)$ and no heat input $(\dot{Q} = 0)$. 
The initial conditions for the resulting ODE's
are obtained from the continuity across $S$ of $J, \: \Pi$ and $F$. 
Thus, $v,\: \rho$, and $T$ for the flow emerging from $S^{+}$ are given by  
\begin{equation}
      \rho v  =  J
\end{equation}
\begin{equation}
      P +  \rho v^{2} = \Pi
\end{equation}
and
\begin{equation}
 J \: \left( \frac{1}{2} v^{2} + \frac{5}{2} a^{2}   \right)  = F  
\end{equation}
where $J,\Pi$, and $F$ are given by Eqs. (13)-(15).

\subsection{Solution branches}

From Eqs. (16)-(18), we readily derive the quadratic equation
\begin{equation}
  v^{2} - 2 u_{p} v + u_{e}^{2} = 0
\end{equation}
where  $u_{p} = 5 \Pi/ 8 J$ and $u_{e} = \surd (F/ 2J)$.
The two solutions are 
\begin{equation}
  v_{\pm} = u_{p} \pm \sqrt{u_{p}^{2} - u_{e}^{2}}
\end{equation}
The corresponding temperatures $T_{\pm}$ are derived from the isothermal sound
speeds given by
\begin{equation}
 a^{2}_{\pm} \: = \: \frac{kT_{\pm}}{\mu m_{H}} \: = 
                              \: u_{e}^{2} - \frac{2}{5} u_{p} v_{\pm}
\end{equation}
and the densities are $\rho_{\pm} = J/ v_{\pm}$.

If $u_{p} = u_{e}$, the two solutions coincide. When this happens,
$v = u_{p} = \surd (5/3) \: a$  - i.e., the outflow at $S^{+}$ is exactly 
sonic.
If $u_{p} > u_{e}$, the solutions are real and distinct. The $v_{+}$
solution is supersonic ($M_{+}$ branch), and the $v_{-}$ solution is
subsonic ($M_{-}$ branch). 
Note that the $M_{+}$ branch has a singularity at $v_{p}/v_{e} = 5/4$,
at which point $a_{+} = 0$.    
Mach numbers for the two branches are plotted 
against $v_{p}/v_{e}$ in Fig. 1.

\begin{figure}
\vspace{8.2cm}
\includegraphics{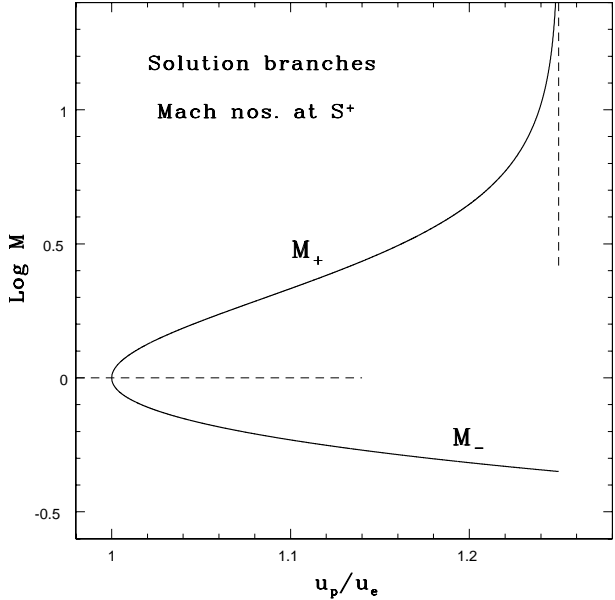}
\caption{Mach number $M$ as a function of $u_{p}/u_{e}$ along the 
supersonic $(M_{+})$ and subsonic $(M_{-})$ solution branches.
When $u_{p}/u_{e} = 5/4$, $M_{+} = \infty$ and
$M_{-} = 1/\surd 5 = 0.447$.}  
\end{figure}

The $M^{-}$ branch corresponds to $S$ being the locus not only of merging 
but also of a stationary shock front. This branch would perhaps be
appropriate if there were a pre-existing slower wind (cf. 
Macfarlane \& Cassinelli 1989), but this is not the circumstance evisaged
here. Instead, therefore, the $M^{+}$ branch is selected since this corresponds
to a high speed two-component flow at $S^{-}$ emerging 
at $S^{+}$ as a single-component supersonic flow.

\subsection{Dissipation at $S$}

In addition to the roots of Eq.(19) being real, a further condition is
mandatory: the transition from $S^{-}$ to $S^{+}$ must be such that 
kinetic energy is
dissipated (entropy production) and not the reverse. For the two branches,
kinetic energy is thermalized at the rates
\begin{equation}
  L_{S}^{\pm} = \frac{1}{2} ( \Phi_{a} v_{a}^{2} + \Phi_{b} v_{b}^{2}
                                               -  \Phi v_{\pm}^{2})
\end{equation}
whose positivity must be checked.

Note that the kinetic energy dissipated at $S$ 
is not radiated away by a thin cooling zone. Instead, this energy
contributes to the flow's enthalpy at $S^{+}$, which then does
$PdV$ work in the subsequent expansion.

\subsection{Outward integration}

The solution for the single-component gas in zone 3 is obtained by 
integrating the equations of motion
\begin{equation}
  v \: \frac{d v}{d r} \: = \: -\frac{1}{\rho} \frac{d P}{d r} - \: g
\end{equation}
continuity 
\begin{equation}
   \Phi = 4 \pi r^{2} \rho v 
\end{equation}
and energy
\begin{equation}
   v \left( \frac{d P}{d r} - \frac{5}{3} \: a^{2}
            \frac{d \rho}{d r} \right) \: = \: -
                  \frac{2}{3} \: n_{e}n_{H} \Lambda(T)
\end{equation}
The initial conditions at $r_{S}$ are $v_{+}, \rho_{+}$ and $T_{+}$ derived
in Sect.4.1.

This integration continues to $r = \infty$. However, this is only possible
if the energy density at $S^{+}$ is sufficient to overcome both the remaining
potential barrier and the cooling losses. If not, a stationary, 
spherically-symmetric wind solution of this type does not exist.

\section{An example}

To illustrate the ideas presented in Sects. 2-4, the solution
for a {\em generic} weak-wind star is now described in detail. 

\subsection{Standard parameters}

The model has several parameters, for which standard 
values are now adopted. Given their uncertainty, sensitivity to
changes are reported in Sect. 6.
  
Because the theory does not predict $\Phi$,
this is derived from previously-tabulated mass fluxes 
(Lucy 2010b; L10b). The chosen model has $T_{\rm eff} = 32.5$kK and 
$\log g = 3.75$, consistent with 
the weak-wind stars $\zeta$ Oph and HD 216532 - see Table 3 in
M09. The model's mass flux 
$J$(gm cm$^{-2}$ s$^{-1}$) = -7.11 dex.

The star's mass ${\cal M} = 24.1 {\cal M}_{\sun}$ is determined by finding the 
point on the ZAMS from which the evolutionary track during core H-burning 
has $\log \: g = 3.75$ when $T_{\rm eff} = 32.5$kK. This point 
is reached after $5.75 \times 10^{6}$ yrs when
$R = 10.83 R_{\sun}$ and the luminosity
$L = 1.18 \times 10^{5} L_{\sun} = 4.52 \times 10^{38}$ erg s$^{-1}$. The
assumed composition is $X=0.70, Z=0.02$.  
 
With $R$ and $L$ determined,  
$\Phi = 4 \pi R^{2} J = 8.80 \times 10^{-9} {\cal M}_{\sun}$ yr$^{-1}$
$ = 1.10 \: L/c^{2}$.
This theoretical $\Phi$ derives from the constraint of regularity at the
sonic point ($v = a$) in the theory of moving reversing layers. In the
weak-wind domain, this theory's predictions exceed the highly uncertain
($\pm 0.7$ dex) observational estimates of M09 by 
$ \approx 0.8$ dex but are lower than the Vink et al. (2000) formula by 
$ \approx  1.4$ dex (Lucy 2010a; L10a).

For the parameters in Eq.(1), we adopt the observationally-supported
O-star values $\beta = 1$ and $v_{\infty} = 2.6 v_{esc}(R) = 2394$ km s$^{-1}$.

The mass $m$ of the blobs must also be specified. Recent
modelling of O-star spectra finds that 'In most cases, clumping must start
deep in the wind, just above the sonic point' (Bouret et al 2008). 
We therefore retain LW's assumption that blobs form at or near the sonic 
point and have diameters comparable to $H_{\rho}$, the local scale height.
At $v=a$ in model $t325g375$, $\rho = 4.27 \times 10^{-14}$ gm cm$^{-3}$ and
$H_{\rho} = 9.86 \times 10^{8}$ cm, so that the crude LW estimate
is $m = 2 \times 10^{13}$ gm.  

The ratio $\eta = \Phi_{b}/\Phi$ must also be specified. 
Following LW, we determine $\eta$ by imposing the constraint that
$\tau_{m} = 1.5$, where    
\begin{equation}
   \tau_{m} =  \min \: [\tau_{1}(r)]  \;\; in \; zone \; 1
\end{equation}
Typically, the minimum occurs at the end of zone 1 where inertial
confinement is greatest. In zone 2, shadowing rapidly becomes irrelevant
since the rapid rise of $T_{a}$ - see Fig.3 - destroys driving ions.

Finally, the conductivity suppression factor $\phi$ introduced in 
Appendix A must be specified. As
standard value, we set $\phi = -1.0$ dex, a moderate degree of suppression
compared to estimates for galaxy clusters (e.g., Ettori \& Fabian 2000).

\subsection{Zone 1}

In this high-density zone close to the photosphere, both components are
assumed to be in thermal equilibrium with the star's radiation field,
a condition approximated by setting $T_{eq} = 0.75 T_{\rm eff}$, as in
L10a,b.  

With the assumptions of isothermal flow, specified $v_{b}$, and no mass
exchange between components, the structure of zone 1 is obtained 
by integrating the ODE
\begin{equation}
   \frac{d \ln v_{a}}{d \ln r} \: = \:  \frac{r}{v_{a}^{2} - a_{a}^{2}}
       \left[\frac{2 a_{a}^{2}}{r} + 
         \frac{\bar{\rho}_{b}}{\bar{\rho}_{a}} \: g_{D} - g  \right]
\end{equation}
The outward integration starts, as in LW, with $v_{b} = 150$ km s$^{-1}$
and $v_{a} = 100$ km s$^{-1}$, a point sufficiently beyond the presumed 
onset of clumpiness that the two-component state may be regarded as
established. The starting radius from Eq.(1) is $r_{i} = 1.067R$.  

Eq.(27) has a singularity when $v_{a} = a_{a}$. Since the integration starts
with $v_{a} > a_{a}$, this singularity only arises if insufficient drag $g_{D}$
causes the flow to decelerate. A parameter set for which this happens
does not admit a steady wind of this type.

As shown in Fig.2, the standard parameters result in an outflow
of ambient gas that accelerates throughout zone 1. This continues until the 
switch to zone 2 is triggered by the onset of the inequality $t_{c} > t_{i}$ 
- see Sect.2.4. 
This occurs at $r/R = 1.28$, with $v_{b} =528$ km s$^{-1}$ and 
$v_{a} =325$ km s$^{-1}$. The relevant time-scales are
$t_{c} = t_{i} =  2.0 \times 10^{3}$ s, which are $\ll$ the local flow 
time-scale, $r/v_{b} = 1.8 \times 10^{4}$ s.  

The post-shock cooling rate $\dot{\cal C}$ required in calculating $t_{c}$
is given by $n_{e} n_{H}\Lambda (T)$, where $\Lambda (T)$ is the optically-
thin cooling function for photospheric abundances tabulated by 
Dere et al. (2009). This rate is computed at the apex of the bow shock with
$n_{e} = 1.18 n_{H}$, corresponding to complete electron-stripping. 
 
At the end of zone 1, the post-shock
temperature has risen to $6.0 \times 10^{5}$K, so X-ray emission
from zone 1 is negligible.

\subsection{Zone 2}

With the isothermal assumption dropped, the structure of zone 2
is determined by Eqs. (4) and (11). With dependent variables $v_{a}$ and
$T_{a}$, the ODE's to be integrated are
\begin{equation}
 (v_{a}^{2} - a_{a}^{2}) \frac{d \ln v_{a}}{d \ln r} \: + \:
                     a_{a}^{2} \frac{d \ln T_{a}}{d \ln r} \: = \:
 2 a_{a}^{2}  + r \left ( \frac{\bar{\rho}_{b}}{\bar{\rho}_{a}} g_{D} - 
                                g  \right)
\end{equation}
and
\begin{equation}
  \frac{2}{3} \: \frac{d \ln v_{a}}{d \ln r} + \frac{d \ln T_{a}}{d \ln r} =  
-\frac{4}{3}+ \frac{2}{3} \frac{r}{P_{a}v_{a}}(\dot{Q} -  \dot{\cal C}_{tot})
\end{equation}
Since all variables are continuous at this transition,
the integration starts at the point 
$(v_{a},v_{b},T_{a},T_{b},r)$ reached by the zone-1 integration. 

Eqs.(28) and (29) are a pair of algebraic equations for the two
derivatives. The determinant of the coefficients' matrix is zero when
$v_{a} = \surd (5/3) a_{a}$ - i.e., at the adiabatic sonic point. 
If this singularity is ecountered, the parameters are inconsistent 
with the conjectured wind structure.

Fig.2 shows that, with the standard parameters,
the flow continues to accelerate throughout zone 2 reaching 
$v_{a} = 940$ km s$^{-1}$ at $r_{S} = 2.14R$, at which point 
$v_{b} = 1277$ km s$^{-1}$.

The corresponding temperature structure predicted for zone 2 is shown in Fig.3. 
At the start, $T_{a,b} = T_{eq} = 24.4$kK. Thereafter, shock-heating
of the ambient component overcomes radiative, conductive and
adiabatic cooling to give a rapidly increasing $T_{a}$, reaching 
the coronal value $10^{6}$K at
$r = 1.35R$ and $3.7 \times 10^{6}$K at $r_{S}$.

The profile for $T_{b}$ shows discontinuous jumps at the beginning and end of 
zone 2.
These result from non-monotonic variations of $\Lambda(T)$. For
example, $\Lambda$'s peak at $T_{\dag}(K) = 5.35$ dex is preceded by 
lower peak at $5.00$ dex. 
Accordingly, after reaching $T_{b}(K) = 5.00$ dex, a slight increase in 
${\cal L}_{in}$
results in a discontinuous jump to $T_{b}(K) = 5.18$ dex, followed quickly by
blob destruction when $T_{b} = T_{\dag}$. Because of these jumps,
the radiative driving of the blobs, which is ultimately reponsible for
$T_{a}$'s increase to coronal values, occurs mostly between 
$T_{b} = 40$ and $90$kK.

Blob temperatures are derived algebraically from Eq.(8) on the assumption that
blobs adjust instantaneously to thermal equilibrium. At $r_{S}$, the
heating time scale 
$1.5 \: nkT_{\dag} \times V_{b}/{\cal L}_{in} = 0.9 \times 10^{2}$ s
compared to the flow time scale $r/v_{b} = 1.3 \times 10^{4}$ s.

In computing cooling rates for blobs, we set $n_{e} = 1.12 n_{H}$, corresponding
to metals being stripped of $\sim 2-3$ electrons.

\begin{figure}
\vspace{8.2cm}
\includegraphics{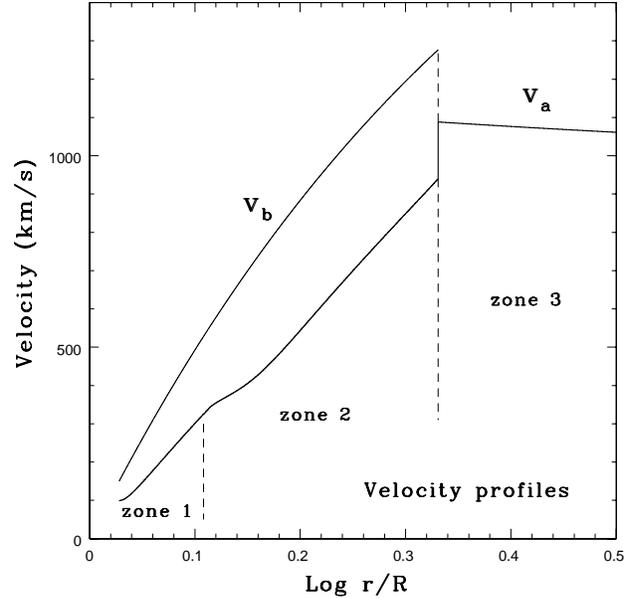}
\caption{Velocities of blobs ($v_{b}$) and ambient gas ($v_{a}$) as functions of
radius. Zone boundaries are indicated. The surface of discontinuity
$S$ where blobs merge with ambient gas occurs at $r/R = 2.14$.}  
\end{figure}
\begin{figure}
\vspace{8.2cm}
\includegraphics{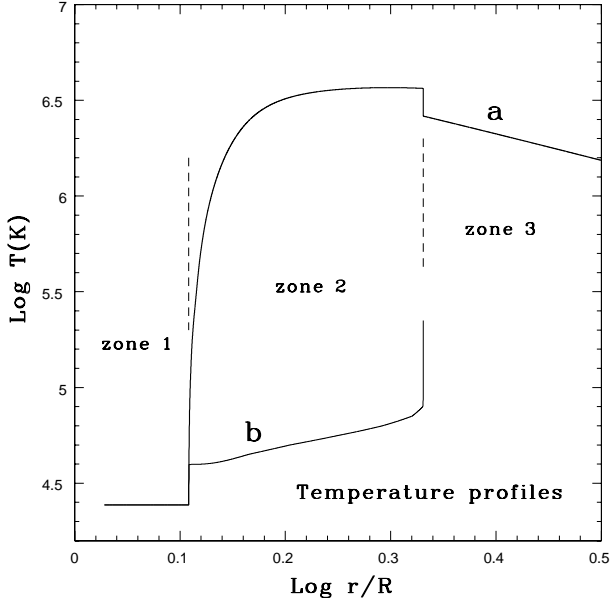}
\caption{Temperatures of blobs ($b$) and ambient gas ($a$) as functions of
radius. Zone boundaries are indicated.}  
\end{figure}

\subsection{Surface of discontinuity S}

At $S^{-}$, the blobs have filling factor $f_{b} = 0.024$, velocity
$v_{b} = 1282$km s$^{-1}$ and temperature $T_{b} = 2.24 \times 10^{5}$K.
The corresponding values for the ambient component are $f_{a} = 0.976$,
$v_{a} = 944$km s$^{-1}$ and $T_{a} = 3.66 \times 10^{6}$K. After merging,
the flow at $S^{+}$ has two possible solutions (Sect. 4.1). 
For the rejected $M_{-}$ solution, the flow
emerges with $v_{-} = 313$km s$^{-1}$, $T_{-} = 1.90 \times 10^{7}$K, 
corresponding to Mach 0.48, and the
implied rate at which kinetic energy is dissipated 
$L_{S}^{-} =  0.31 \times 10^{34}$ erg s$^{-1}$ or $6.9 \times 10^{-6} \:L$. 

 For the selected $M_{+}$ solution, the flow
emerges with $v_{+} = 1094$km s$^{-1}$, $T_{+} = 2.58 \times 10^{6}$K, 
corresponding to Mach 4.6, and the
implied dissipation rate  
$L_{S}^{+} =  0.92 \times 10^{32}$ erg s$^{-1}$ or $2.0 \times 10^{-7} \:L$. 

 Notice that $v_{+} \in (v_{a}, v_{b})$, as expected if $S$ is the locus
only of merging. In contrast,
$v_{-} < v_{a}$, so there is a coincident shock, as also indicated
by the far greater dissipation rate $L_{S}^{-}$.

\subsection{Zone 3}

The single component flow emerging from $S$ is a pure coronal wind: the only
outward force is the gradient of thermal pressure.

The structure of zone 3 is obtained by continuing the integration of 
Eqs.(28) and (29), but now with $g_{D} = 0, \; \dot{Q} = 0$ and
$\: \dot{\cal C}_{tot} = n_{e}n_{H}\Lambda (T)$. The initial conditions
at $r_{S}$ are $v_{+}$ and $T_{+}$ given in Sect. 5.4.

A short segment of this outflow is plotted in Figs. 2 and 3, showing that 
the flow decelerates and (inevitably) cools. For these standard parameters,
the energy density at $S$ suffices to overcome cooling and power  
escape to $\infty$. At $r_{f}/R =100$, the flow has slowed to $984$km s$^{-1}$,
way beyond the local $v_{esc} = 92$km s$^{-1}$ 

The temperature drops below the coronal value $10^{6}$K
at $r/R = 4.24$ and to $10^{5}$K at $r/R = 13.2$.

\subsection{Emission measure}

With standard parameters, our generic weak-wind star is predicted to have a 
corona ($T > 10^{6}$) that extends from $r_{1} = 1.35R$ to $r_{2} = 4.24R$ 
and so will be an X-ray emitter. As
a crude guide to detectability, we compute the
emission measure of coronal gas
\begin{equation}
  \varepsilon \: = \:4 \pi \int_{r_{1}}^{r_{2}}   n_{e}n_{H} \: r^{2} dr
\end{equation}
and its hardness parameter
\begin{equation}
  <kT> \: = \: 4 \pi \varepsilon^{-1} 
                     \int_{r_{1}}^{r_{2}} kT \: n_{e}n_{H} \: r^{2} dr 
\end{equation}
The results are $\varepsilon(cm^{-3}) = 53.51$ dex and 
$<kT> = 0.20$ keV.

\subsection{Energy budget}

The global energy budget of this multi-zone wind
is of interest. The input is the rate of working in zones 1 and 2 of $g_R$, 
the force per unit mass acting on the blobs. This rate  
$L_{wrk} = 5.4 \times 10^{33}$ erg s$^{-1}$. 

The balancing output is $L_{M}+L_{W}$, where $L_{M}$ is the rate at
which matter gains kinetic and potential energy, and
$L_{W}$  is the wind's radiative luminosity.
For the interval $(r_{i},r_{f})$, $L_{M} = 4.8\times 10^{33}$ erg s$^{-1}$
or 88.5\% of $L_{wrk}$. The remaining 11.5\% is accounted for by  
$L_{W}$, which comprises radiative losses from shock fronts in
zone 1, cooling radiation from blobs and ambient gas in zone 2, and
cooling radiation from the coronal flow in zone 3.

For an idealized line-driven wind in which gas remains (by assumption) 
at $T_{eq}$, $PdV$ work is negligble so that $L_{M} = L_{wrk}$. In contrast, 
for a pure coronal wind, $L_{wrk} = 0$, so that $L_{M}$ is entirely due
to the $PdV$ work of the hot gas. The relative contributions of these two
mechanisms in this hybrid case is of interest.

In answering this, we must first integrate $\dot{Q}$ from Eq.(9) over 
zones 1 and 2 to obtain the total dissipation rate 
$L_{D} = 1.1 \times 10^{33}$ erg s$^{-1}$. The quantity 
$L_{wrk} - L_{D} = 4.3 \times 10^{33}$ erg s$^{-1}$ is then the contribution
to $L_{M}$ due {\em directly} to radiative driving. On the other hand,
the contribution of $PdV$ work by hot gas is    
$L_{D} - L_{W} = 0.4 \times 10^{33}$ erg s$^{-1}$. 

A measure of the proximity of a hybrid- to a pure coronal wind is the ratio   
\begin{equation}
   \theta = (L_{D} - L_{W})/L_{M}
\end{equation}
which $= 0$ for a conventional line-driven wind and $= 1$ for a coronal wind.
With standard parameters, the multi-zone wind has $\theta = 0.08$, so
direct radiative driving still dominates in accounting for $L_{M}$. 

A further quantity of interest is the integrated cooling rate 
of gas with $T_{e} > 10^{6}$K, since this is approximately the 
wind's X-ray luminosity. For zones 2 and 3, this gives 
$L_{X} \approx 3.4 \times 10^{31}$ erg s$^{-1}$,
so that $L_{X}/L  \approx  0.76 \times 10^{-7} L$, similar to the ratio found 
for early-type O stars.

\section{Non-standard parameters}

The theory developed in Sects. 2-4 has several parameters,
each of which would either be predicted or rendered unnecessary if calculations
could be carried out from first principles. Sensitivity of the results to 
these currently unavoidable parameters must therefore be investigated. 
Accordingly, sequences of solutions are now reported in which a single 
parameter is varied while keeping others at the standard values of Sect. 5.

Key properties of the models are given in Table 1. The 
quantities reported are as follows:\\

{\em Col. 1}: Sequence identifier.\\

{\em Col. 2}: Exponent in Eq.(1), the velocity law.\\

{\em Col. 3}: Log of total mass-loss rate in ${\cal M}_{\sun}$ yr$^{-1}$.\\

{\em Col. 4}: Log of blobs' mass in gm.\\

{\em Col. 5}: Log of conductivity suppression factor - see Eq. (A.9).\\

{\em Col. 6}: Fraction of mass-loss in blobs = $\Phi_{b}/\Phi$ .\\

{\em Col. 7}: Shadowing optical depth - see Eq.(26).\\

{\em Col. 8}: $v_{b}$ in km s$^{-1}$ at the destruction radius $r_{S}$.\\

{\em Col. 9}: Maximum ambient gas temperate in $10^{6}$K.\\

{\em Col. 10}: Log of emission measure in cm$^{-3}$ - see Eq.(30).\\

{\em Col. 11}: Hardness parameter in keV - see Eq.(31).\\

\begin{table*}

\begin{minipage}{80mm}

\caption{Solutions with non-standard parameters.}

\label{table:1}

\centering

\begin{tabular}{c c c c c c c c c c c}

\hline\hline

 Seq.   & $\beta$ & $\log \Phi$ & $\log m$ & $\log \phi$ & $\eta$ & $\tau_{m}$ &
 $v_{b}(r_{S})$ & $T_{max}$ & $\log \varepsilon$ & $<kT> $  \\

\hline
\hline

  I & 1.0 &   -8.06 &   13.3  &  -3.0 &   0.45 &   1.5 &    2286 &     8.4 &   53.43 & 0.40 \\
    & 1.0 &   -8.06 &   13.3  &  -2.5 &   0.45 &   1.5 &    2180 &     8.2 &   53.44 & 0.39 \\
    & 1.0 &   -8.06 &   13.3  &  -2.0 &   0.45 &   1.5 &    1991 &     7.7 &   53.44 & 0.36 \\
    & 1.0 &   -8.06 &   13.3  &  -1.5 &   0.45 &   1.5 &    1678 &     6.2 &   53.47 & 0.30 \\
    & 1.0 &   -8.06 &   13.3  &  -1.0 &   0.45 &   1.5 &    1277 &     3.7 &   53.51 & 0.20 \\
    & 1.0 &   -8.06 &   13.3  &  -0.5 &   0.45 &   1.5 &    1015 &     2.1 &   53.47 & 0.13 \\ 
    & 1.0 &   -8.06 &   13.3  &   0.0 &   0.45 &   1.5 &     873 &     1.4 &   53.18 & 0.11 \\

\cline{1-11}

 II & 1.0 &   -8.26  &  13.3  &  -1.0  &  0.47 &   1.5 &    1109  &    3.5  &  53.27 & 0.19 \\
    & 1.0 &   -8.06  &  13.3  &  -1.0  &  0.45 &   1.5 &    1277  &    3.7  &  53.51 & 0.20 \\
    & 1.0 &   -7.76  &  13.3  &  -1.0  &  0.41 &   1.5 &    1509  &    3.7  &  53.92 & 0.21 \\
    & 1.0 &   -7.46  &  13.3  &  -1.0  &  0.37 &   1.5 &    1719  &    3.6  &  54.31 & 0.21 \\
    & 1.0 &   -7.16  &  13.3  &  -1.0  &  0.32 &   1.5 &    1918  &    3.3  &  54.70 & 0.21 \\
    & 1.0 &   -6.86  &  13.3  &  -1.0  &  0.27 &   1.5 &    2121  &    2.9  &  55.06 & 0.19 \\

\cline{1-11}

III & 0.5  &   -8.06  &  13.3  &  -1.0  &  0.68  &  1.5  &   1348  &   3.6  &  53.14 & 0.17 \\
    & 0.6  &   -8.06  &  13.3  &  -1.0  &  0.59  &  1.5  &   1380  &   4.0  &  53.37 & 0.20 \\
    & 0.8  &   -8.06  &  13.3  &  -1.0  &  0.50  &  1.5  &   1341  &   3.9  &  53.49 & 0.20 \\
    & 1.0  &   -8.06  &  13.3  &  -1.0  &  0.45  &  1.5  &   1277  &   3.7  &  53.51 & 0.20 \\
    & 1.5  &   -8.06  &  13.3  &  -1.0  &  0.37  &  1.5  &   1135  &   3.2  &  53.50 & 0.17 \\
    & 2.0  &   -8.06  &  13.3  &  -1.0  &  0.32  &  1.5  &   1033  &   2.8  &  53.45 & 0.15 \\
    & 2.5  &   -8.06  &  13.3  &  -1.0  &  0.29  &  1.5  &    962  &   2.5  &  53.40 & 0.14 \\

\cline{1-11}

 IV & 1.0  &   -8.06  &  11.9  &  -1.0  &  0.20  &  1.5  &    889  &   2.6  &  53.81 & 0.14 \\
    & 1.0  &   -8.06  &  12.3  &  -1.0  &  0.26  &  1.5  &    989  &   3.0  &  53.74 & 0.16 \\
    & 1.0  &   -8.06  &  12.8  &  -1.0  &  0.35  &  1.5  &   1135  &   3.4  &  53.63 & 0.18 \\
    & 1.0  &   -8.06  &  13.3  &  -1.0  &  0.45  &  1.5  &   1277  &   3.7  &  53.51 & 0.20 \\
    & 1.0  &   -8.06  &  13.8  &  -1.0  &  0.55  &  1.5  &   1400  &   4.0  &  53.38 & 0.21 \\
    & 1.0  &   -8.06  &  14.3  &  -1.0  &  0.66  &  1.5  &   1503  &   4.2  &  53.23 & 0.22 \\
    & 1.0  &   -8.06  &  14.6  &  -1.0  &  0.72  &  1.5  &   1555  &   4.4  &  53.12 & 0.22 \\

\cline{1-11}

  V & 1.0  &   -8.06  &  13.3  &  -1.0  &  0.41  &  1.0  &   1322  &   4.2  &  53.67 & 0.24 \\
    & 1.0  &   -8.06  &  13.3  &  -1.0  &  0.45  &  1.5  &   1277  &   3.7  &  53.51 & 0.20 \\
    & 1.0  &   -8.06  &  13.3  &  -1.0  &  0.48  &  2.0  &   1246  &   3.3  &  53.40 & 0.17 \\
    & 1.0  &   -8.06  &  13.3  &  -1.0  &  0.50  &  2.5  &   1225  &   2.9  &  53.31 & 0.16 \\

\hline
\hline

\end{tabular}

\end{minipage}

\end{table*}

\begin{figure}
\vspace{8.2cm}
\includegraphics{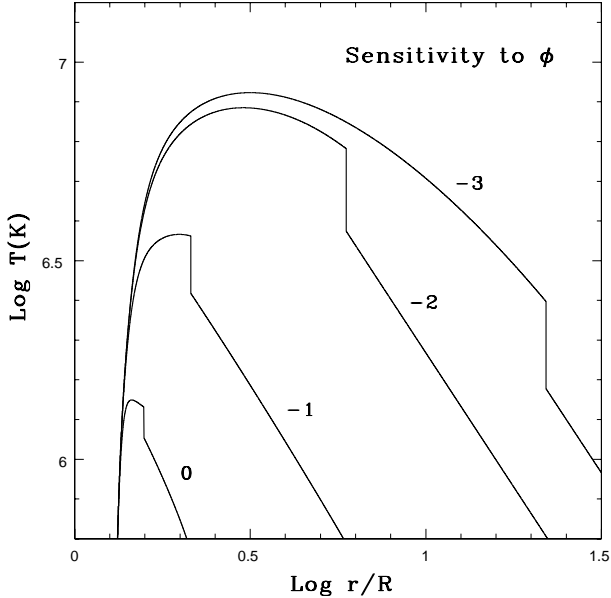}
\caption{Sensitivity of coronal winds to $\phi$, the magnetic suppression
factor - see Eq.(A.9). Values of $\log \phi$ are shown. The vertical segments
are the surfaces of discontinuity $S$.}  
\end{figure}

\subsection{Sequence I}

In this sequence, the conductivity suppression factor varies from
$\phi = 0.001$, an extreme value but with observational support for galaxy 
clusters (Ettori \& Fabian 2000), to $\phi =1$, the value 
for a non-magnetized plasma.

Not surprisingly, the predictions are highly sensitive to $\phi$, 
and this might eventually be exploited diagnostically. With $\phi = 1$,
conductive heating of the blobs destroys them already at $r_{S} = 1.57R$ where
$v_{b} = 873$km s$^{-1}$. 
However, with $\phi = 0.001$, blobs survive out to
$r_{S} = 22.0R$ where $v_{b} = 2286$km s$^{-1}$.

Table 1 also shows that coronal temperature and the hardness parameter
increase as $\phi \rightarrow 0$. However, the emission measure 
$\varepsilon$ remains $\sim 53.4 - 53.5$ dex after an initial sharp rise
from $53.18$ dex for $\phi = 1$. The predicted temperature profiles of
the coronae as $\phi$ varies are plotted in Fig.4.

This sequence demonstrates the diagnostic potential of UV and X-ray
data in constraining magnetic suppression of conductivity. The UV data
measures the highest velocity at which wind matter transfers photon
momentum to the gas and the X-ray data measures the hardness of coronal
emission. 

A further diagnostic test provided by P Cygni lines is the weakness
of emission components. As $r_{S}$ decreases with increasing $\phi$,
the fraction of scattered photons occulted by the star increases and the
emission component weakens. This effect was invoked for $\tau$ Sco by LW
in arguing that 'outflowing gas loses its ability to scatter UV radiation 
while still close to the star's surface.' Note that the weak-wind stars 
investigated in M09 all have C {\sc iv} resonance doublets
with weak or absent emission components. Diagnostic modelling of these stars
would improve if UV scattering were truncated at finite radius.

\subsection{Sequence II}

As noted in Sect.5.1, the $\Phi$ of weak-wind stars is poorly determined.
This sequence explores sensitivity to this uncertain parameter.  

When $\Phi$ is increased above the standard value from L10b, the blobs 
survive to higher velocities, and the higher coronal densities give
the approximate scaling law $\varepsilon \propto \Phi^{1.3}$. Interestingly,
the quantities $T_{max}$ and $<kT>$ are insensitive to $\Phi$.

The attempt to continue this sequence to lower $\Phi$'s failed at $-8.36$ dex
because the singularity in zone 2 discussed in Sect.5.3 is encountered.
This arises as follows: 
the sharp initial rise of $T_{a}$ in zone 2 causes $M$ to decrease despite 
increasing $v_{a}$ - see Figs.2 and 3. But as
$T_{a}$ levels off $M$ reaches a minimum and then rises again. Sequence II
terminates for $\Phi$ between $-8.36$ and $-8.26$ dex when this minimum falls 
to $M = 1$. For the solution plotted in Figs. 2 and 3, this zone-2 minimum
is $M = 1.88$ at $r = 1.47R$. 

\subsection{Sequences III-V}

In sequence III, the velocity-law exponent varies from $\beta = 0.5$ - 
rapid acceleration -
to  $\beta = 2.5$ - slow acceleration. The standard value $\beta = 1.0$ is 
approximately a stationary point
as regards the coronal properties $\varepsilon$ and $<kT>$, so these are 
insensitive to $\beta$. 
However, $v_{b}(r_{S})$ is moderately sensitive.   

In sequence IV, solution sensitivity to the highly uncertain blob mass is 
explored. Fortunately, coronal properties are only moderately sensitive,
with $\varepsilon \propto m^{-0.22}$ and $<kT> \: \propto m^{0.03}$. 
In regard to blob destruction, this occurs as expected
at low velocities for small $m$. In consequence, sequence IV terminates
for $m(gm)$ between 11.8 and 11.9 dex because the outflow in zone 3
is then unable to reach $\infty$ on account of negative energy density -
see Sect.4.3.  

Finally, sensitivity to the shadowing parameter $\tau_{m}$ is investigated 
with sequence V. Again only moderate sensitivity is found.

\section{Conclusion}

The aim of this paper has been to investigate the structural changes
of O-star winds when $\Phi$ decreases to the extent found for the 
{\em weak-wind} stars.
To this end, the two-component phenomenological model developed originally 
for $\zeta$ Puppis 
is modified to incorporate LW's conjectures following
the breakdown of that model's assumptions for 
$\tau$ Sco. 
When applied to a generic weak-wind star, the revised model predicts that 
shock-heating of the ambient gas
gives rise to coronal temperatures, that conductive-heating eventually 
destroys the blobs, and that the resulting single-component flow
coasts to $\infty$ as a pure coronal wind. Thus, in broad outline,
the volumetric roles of hot and cool gas in O-star winds are reversed. In 
the now standard picture for a star such as
$\zeta$ Puppis the  X-ray emitting gas occupies a tiny fraction of 
the wind's volume,
with the bulk of the volume being highly-clumped cool gas with $T \sim T_{eq}$.
In contrast, in the picture suggested here for the weak-wind stars, 
X-ray emitting gas fills most of the volume for $r \ga 1.3 R$, with surviving 
cool gas in the form of dense clumps with $f_{b} \sim 0.01 - 0.03$. 

As is common elsewhere in astrophysics, the approach adopted in this paper
is phenomenological modelling. 
A simplified picture of the phenomenon is combined 
with approximate treatments of the 
expected physical effects to create an 'end-to-end' tractable theory
that obeys conservation laws and 
makes testable predictions.
Such theories are
of course always an interim measure, to be discarded when the obstacles to 
calculation from first principles are overcome. Unfortunately, in this case, 
these obstacles are formidable: 3-D time-dependent gas dynamics, radiative
transfer, and heat conduction including saturation and possibly magnetic 
suppression.

Evidently, the fundamental approach is unlikely to yield results anytime soon. 
Accordingly, possible improvements of the crude modelling described
herein should be investigated. Also
diagnostic codes should incorporate features of such models to extract
more reliable parameters from observational data.  

\appendix

\section{Heat conduction into a spherical blob}

In zone 2, the blobs are surrounded by gas whose temperature
is rising to coronal values. Conduction will therefore transfer heat
into the blobs, and this constitutes a loss term in the energy
equation for the ambient gas.

Given that $f_{b} \ll 1$, it suffices to consider a single spherical 
blob with temperature $T_{b}$ and radius $\sigma$ located  
at $r = 0$ in an infinite medium with $T \rightarrow T_{a}$ as
$r  \rightarrow \infty$. If $\dot{\cal C}$ 
is the cooling rate per unit volume and $\kappa$ is the conductivity, 
the equilibrium temperature profile for $r > \sigma$ is given by the 
equations 
\begin{equation}
  \frac{d {\cal L}}{d r} = 4 \pi r^{2} \: 
                      [\: \dot{\cal C} (T) -\dot{\cal C} (T_a) \:] 
\end{equation}
and   
\begin{equation}
  \frac{d T}{d r} = \frac{{\cal L}}{4 \pi r^{2} \kappa}
\end{equation}
with boundary conditions
\begin{equation}
  T(\sigma) = T_{b}    \:\;\;\; and \;\;\;\;  T(\infty) = T_{a}
\end{equation}
In the blob's absence, the gas is isothermal and has cooling rate 
$\dot{\cal C} (T_{a})$, which is 
subtracted in Eq. (A.1). Accordingly, 
${\cal L} (\infty)$ 
is the {\em additional} cooling due to the blob's presence. Of this, 
$\Delta {\cal L} = {\cal L} (\infty) -  {\cal L} (\sigma)$ represents
emission from ambient gas cooled below $T_{a}$ by the blob, 
and ${\cal L} (\sigma)$ is the rate of heat conduction into the blob.

In thermal equilibrium, ${\cal L} (\sigma)$ 
is balanced by emission from within the blob. Now, since radiative
cooling is $\propto \rho^{2}$ and $\rho_{b} \gg \rho_{a}$, we expect that
$\Delta {\cal L} \ll {\cal L} (\sigma)$. Therefore,
to a first approximation, ${\cal L} (r > \sigma) =  {\cal L} (\sigma)$,
a constant, and this allows Eq. (A.2) to be solved analytically when
$\kappa \propto T^{5/2}$ (Spitzer 1962). The resulting temperature
profile is given by

\begin{equation}
  t^{7/2} = t^{7/2}_{b} + (1- t^{7/2}_{b})(1-\frac{\sigma}{r})
\end{equation}
where $t = T(r)/T_{a}$, and the corresponding rate of heat conduction into 
the blob is
\begin{equation}
  {\cal L}_{cl} =  \frac{8}{7}\:\pi \sigma \:
                                    [\:(\kappa T)_{a} -(\kappa T)_{b}\:]
\end{equation}
Since $\kappa T \propto T^{7/2}$, ${\cal L}_{cl}$ is insensitive to $T_{b}$
when $T_{a} \gg T_{b}$. For the solutions reported in Sects. 5 and 6, 
we take $\kappa = 1.0 \times 10^{-6} T^{5/2}$, corresponding to Coulomb
logarithm $\ln \Lambda = 17$. 

The above discussion treats conduction in the diffusion limit - i.e.,
where the mean free path of the electrons is $\ll$ macroscopic length scales. 
In the
opposite limit, heat conduction into the blob is flux-limited and saturates at
\begin{equation}
  {\cal L}_{sat} =  4 \pi \sigma^{2} \: q_{sat}
\end{equation}
where $q_{sat}$ is estimated by Cowie \& McKee (1977)
to be 
\begin{equation}
  q_{sat} = 0.4 \: \left( \frac{2kT_{e}}{\pi m_{e}} \right)^{1/2} n_{e}kT_{e}
\end{equation}
and is here evaluated at $T_{a}, (n_{e})_{a}$.
Interpolating between these limits
(cf. Balbus \& McKee 1982), we take the rate of heat conduction into the blob
to be ${\cal L}_{cond}$, where
\begin{equation}
   {\cal L}^{-1}_{cond} =  {\cal L}_{cl}^{-1} + {\cal L}_{sat}^{-1}
\end{equation}
At high temperatures, this gives ${\cal L}_{cond} \propto T^{3/2}$ in place of
${\cal L}_{cl} \propto T^{7/2}$.

The above formula is for a non-magnetized plasma.
But since stars in and near the weak-wind domain have detected magnetic
fields (e.g., Oskinova et al. 2011), we include the possibility of
magnetic suppression of heat conduction by writing
\begin{equation}
   {\cal L}_{in} =  \phi \: {\cal L}_{cond}
\end{equation}
In this investigation, $\phi$ is varied to explore its impact
on the solutions.
In future, it may be determined or constrained by fitting
observational data. 

The suppression of thermal conductivity in astrophysical plasmas has been
strikingly confirmed by
the discovery of cold fronts in X-ray maps of clusters of galaxies  
(e.g., Carilli \& Taylor 2002). For the cluster Abell 2142, 
Ettori \& Fabian (2000) estimate a reduction factor of between 250 and 2500. 
They speculate that, as a result of merging, different magnetic structures
are in contact and so remain to high degree thermally isolated. The 
displacements of wind clumps from their nascent ambient surroundings might
well lead similarly to substantial reduction factors.

\end{document}